\documentclass[a4paper,11pt]{article}
\pdfoutput=1 

\usepackage{jinstpub} 

\title{\boldmath A Timing RPC with low resistive ceramic electrodes}

\author[a]{R. Sultanov,}
\author[a]{A. Akindinov,}
\author[b]{R. Beyer,}
\author[b]{J. Dreyer,}
\author[b]{X. Fan,}
\author[b]{R. Greifenhagen,}
\author[b]{B. K\"ampfer,}
\author[b]{R. Kotte,}
\author[b]{A. Laso Garcia,}
\author[a]{D. Malkevich,}
\author[b,1]{L. Naumann,\note{Corresponding author.}}
\author[a]{V. Plotnikov,}
\author[a]{M. Prokudin,}
\author[a]{S. Shirinkin,}
\author[b]{and D. Stach.}

\affiliation[a]{Institute for Theoretical and Experimental Physics, Moscow, Russia}
\affiliation[b]{Helmholtz-Zentrum Dresden-Rossendorf, Dresden, Germany}

\emailAdd{L.Naumann@hzdr.de}

\abstract{For precise start time determination a Beam Fragmentation T$_0$ Counter (BFTC) is under development
for the Time-of-Flight Wall of the Compressed Baryonic Matter Spectrometer (CBM) at the 
Facility for Antiproton and Ion Research (FAIR) at Darmstadt/Germany. 
This detector will be located around the beam pipe, covering the front area  of the Projectile 
Spectator Detector. The fluxes at this region are expected to exceed 10$^5$cm$^{-2}$s$^{-1}$. 
Resistive plate chambers (RPC) with ceramic composite electrodes  could be use because of their 
high rate capabilities and radiation hardness of material. Efficiency $\ge$\,97\,\%, 
time resolution $\le$\,90\,ps and rate capability 
$\ge$\,10$^5$cm$^{-2}$s$^{-1}$ were confirmed during many tests with high beam fluxes of relativistic 
electrons. We confirm the stability of these characteristics 
with low resistive Si$_3$N$_4$/SiC floating electrodes for a prototype of eight small RPCs, where each of 
them contains six gas gaps. The active RPC size amounts 20$\times$20\,mm$^2$ produced on basis of Al$_3$O$_2$ 
and Si$_3$N$_4$/SiC ceramics.  Recent test  results obtained with relativistic electrons at the linear 
accelerator ELBE of the Helmholtz-Zentrum Dresden-Rossendorf with new PADI-10 Front-end electronic will 
be presented.
}


\collaboration[c]{on behalf of CBM-TOF collaboration}

\proceeding{XIV$^{\text{th}}$Workshop on Resistive Plate Chambers and related detectors\\
  19. - 23. Feb. 2018\\
  Mexico, Puerto Vallarta}

\begin{document}
\maketitle
\flushbottom

\section{Introduction}
\label{sec:intro}

Within the framework of the Facility for Antiproton and Ion Research, which is currently 
under construction at Darmstadt/Germany, a determined effort is being made to implement the CBM 
spectrometer ~\cite{Senger2002}. Important prerequisites of high energy heavy ion experiments are 
the start-time and the reaction-plane determination. 
For the CBM Time-of-Flight Wall the use of resistive plate chambers (RPC) for the 
Beam Fragmentation T$_0$ Counter (BFTC) with low resistive radiation 
hard ceramics electrodes and small chess-board like 
single cells is under consideration ~\cite{Akindinov2017, Deppner2014}. This detector should 
cover the inner solid angle range of the spectrometer with an area
of about 120$\times$120\,cm$^2$ in front of the Projectile Spectator Detector. 
It is planned to arrange approximately 4000 RPC cells of 20$\times$20\,mm$^2$ size around the beam 
tube with a central hole of 40$\times$40\,cm$^2$ to detect the arrival time and the position of the charged 
relativistic beam fragments and hence, to determine the reaction time and reaction plane of an event. 
Due to the expected high fluxes, the detector is fixed in terms of both efficiency and time resolution. 
The small size of the detector cells is mandatory to minimize the  occupancy in a single cell guaranteeing a double hit
probability of less than 2\,\% ~\cite{Sultanov2013}. Also the crosstalk between adjacent detector cells should not exceed 2\,\%.
Tests with the BFTC-minimodule of 8 RPC cells have been performed ~\cite{Akindinov2017, Akindinov2017a} 
with high fluxes of  electrons at the ELBE accelerator during two beam-times in 2017 in order to adopt 
the BFTC-prototype construction and signal readout scheme for operation with the PADI ASIC ~\cite{PADI}.
RPCs with electrode areas from 20$\times$20\,mm$^2$ up to 200$\times$200\,mm$^2$  have been assembled and tested
with electrons and protons under high irradiation conditions ~\cite{Naumann2011, LasoGarcia2016}. 
The low-resistive electrodes have been exposed with neutrons to estimate the maximum of non-ionizing  radiation doses 
without a modification of the ceramics properties ~\cite{LasoGarcia2012, Arefiev1996}.

\section{Instrumentation}
\label{sec:intro}

\begin{figure}[htb]
\centering
\includegraphics*[width=120mm]{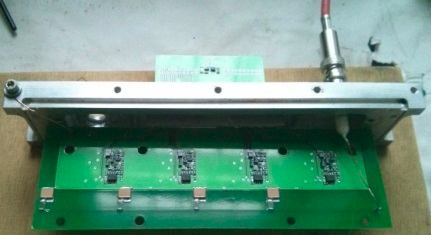} 
\includegraphics*[width=122mm]{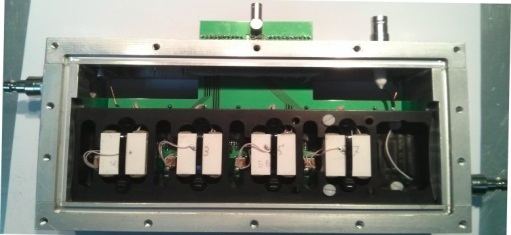} 
\caption{\label{fig:1}Top - The PCB readout board with HV distribution (HV capacitors) and 
preamplifier/shaper circuits fixed on the upper cover of the detector box; Bottom -  
RPCs of the prototype inside the box. On the photo four of eight RPCs (white quadrat with black belt) are shown. 
Four RPCs are located on the back side. All RPCs are sorted meandering along a chain.}
\label{l2ea4-f1}
\end{figure}

The BFTC-minimodule represents a gas volume with a support structure for 8 RPCs inside, which consists of a 
plastic (POM) frame with assembled RPCs on it. To omit geometrically acceptance losses, the RPCs are sorted  
meandering. Each quadratic RPC with 6 gas gaps of 250\,$\mu$m consists of a stack of three separate 
counter cells ~\cite{Akindinov2017, Akindinov2017a}. 
The counter gas  mixture consists of two components, R134a\,(90\,\%) and SF$_6$\,(10\,\%) where the detector 
volume has been exchanged twice per hour.
The central electrodes at floating potential are made of low resistive Si$_3$N$_4$/SiC ceramic composite, 
while all anodes and cathodes are metallic electrodes on a
high resistive Al$_2$O$_3$ backing.  To reduce the dark current of the detector, all electrode edges are 
grooved with Rogowski shapes ~\cite{Sultanov2015}. 
The RPCs are fixed on a PCB board inside the detector box which serves the following functions: 
high voltage distribution for all 8 RPCs from a single input and primary readout and converting 
the RPC signal from single ended to differential mode (100\,$\Omega$) for signal transmission 
to the PADI-10. The  signal amplification amounts to 1.5 and shaping of the signal is performed in the readout 
scheme, so a typical output pulse has no overshoot and its length amounts to 5.5\,ns at 
the base level. This ensures a readout without pileup at high counting rates, expected 
at the BFTC region ~\cite{Sultanov2013}. The current design of the board assures that 
the high voltage lines are placed in the inner layer of the PCB, to minimizes the cross-talks via the HV line. 
The board with the upper cover of the gas tight box and with assembled RPCs is shown in 
figure~\ref{fig:1}.  

The detector was tested with a high flux of single relativistic electrons of energy\,$\ge$\,30\,MeV with 
a dedicated test facility at ELBE with a scintillating counter telescope and additional trigger and timing information 
from the accelerator ~\cite{LasoGarcia2016, Naumann2010, Naumann2011, ELBE2000}. 
As readout electronics for the RPC signals PADI (versions 6 and 10 at first and second 
beam-time, correspondingly), ASIC and VFTX TDC were used, while a CAEN TDC was used for the
trigger scintillators. A trigger logic was realized in CAEN FPGA module V1495. 
Since the VFTX TDCs were working with external clock, the accelerator RF signal was 
fed into the second VFTX TDC. 

\section{Results}
\label{sec:intro}
\begin{figure}[htb]
\centering
\includegraphics*[width=75mm]{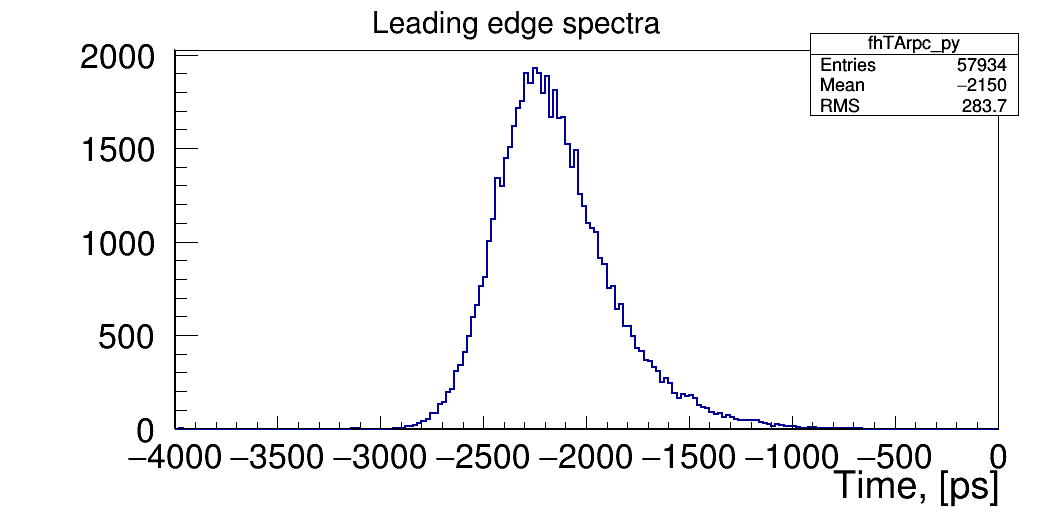} 
\includegraphics*[width=75mm]{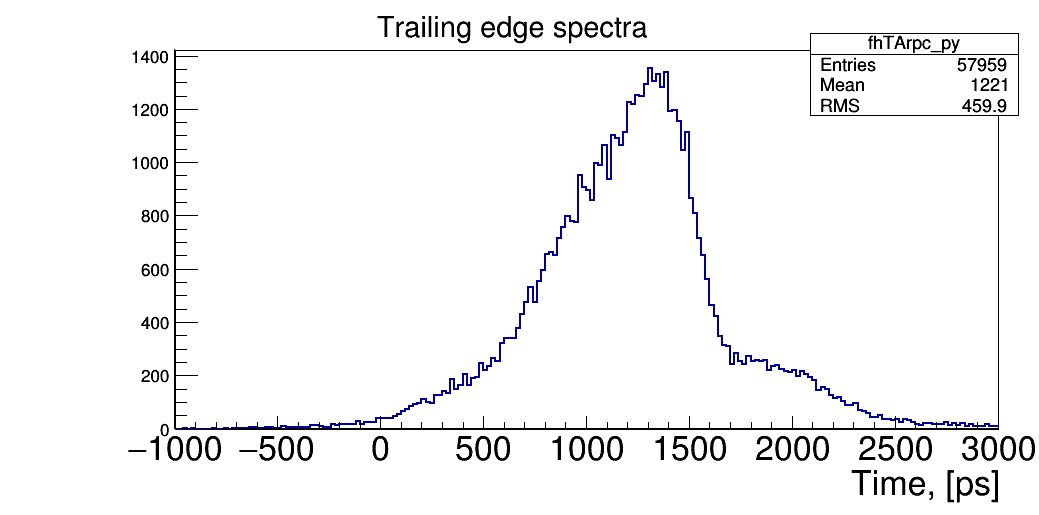} 
\includegraphics*[width=75mm]{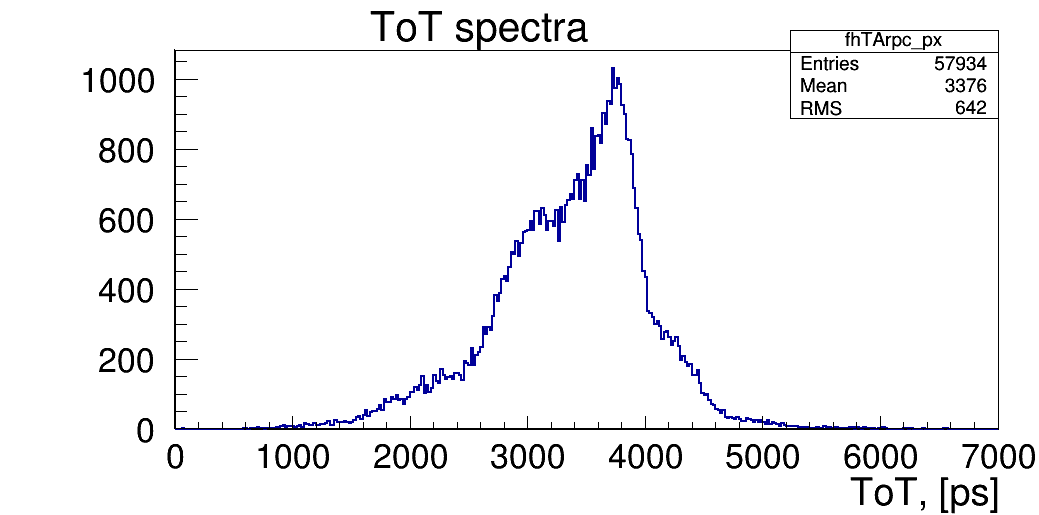} 
\includegraphics*[width=75mm]{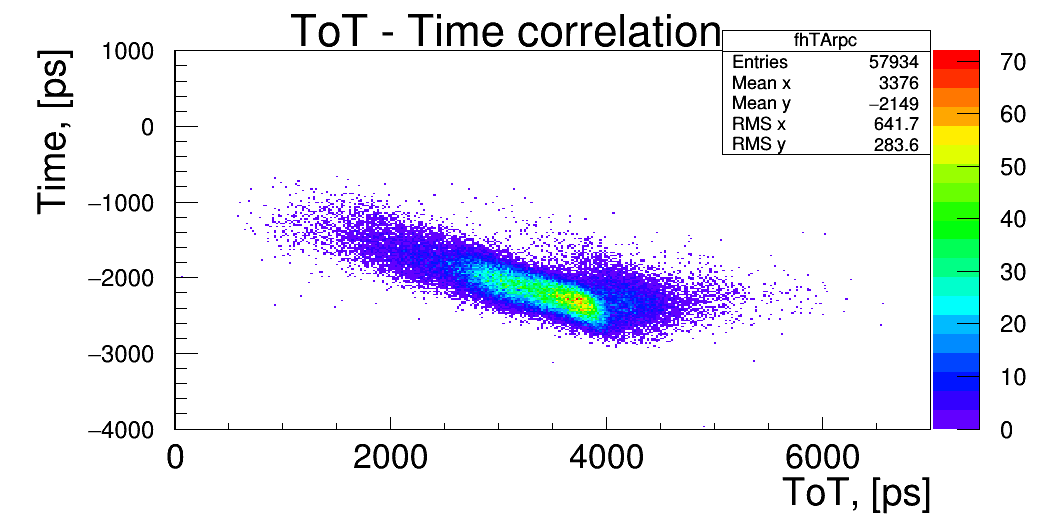} 
\caption{\label{fig:2}Spectra from top-left to bottom-right: leading edge distribution, trailing edge 
distribution, Time-over-Threshold (ToT) distribution, ToT-to-time correlation.}
\label{l2ea4-f2}
\end{figure}

The arrival time of the RPC signal 
is  determined as difference between leading edge of the RPC signal and the reference signal. 
Characteristic time  spectra for the leading and trailing edge distributions and for the amplitude
dependend Time-over-Threshold (ToT) behavior are shown in figure~\ref{fig:2} for the RPC cell \#\,4 at
92\,kV/cm. 
The leading edge distribution shows a steep rise and the trailing edge distribution  a steep fall of 
the time signal, while  the ToT spectra exhibits a double peak structure. The correlation between ToT and timing
allows for a walk correction of the timing distribution.
The optimal value of the signal amplitude threshold at the PADI ASIC input was determined by a noise scan 
and amounts 180\,mV. 

\begin{figure}[htb]
\centering
\includegraphics*[width=120mm]{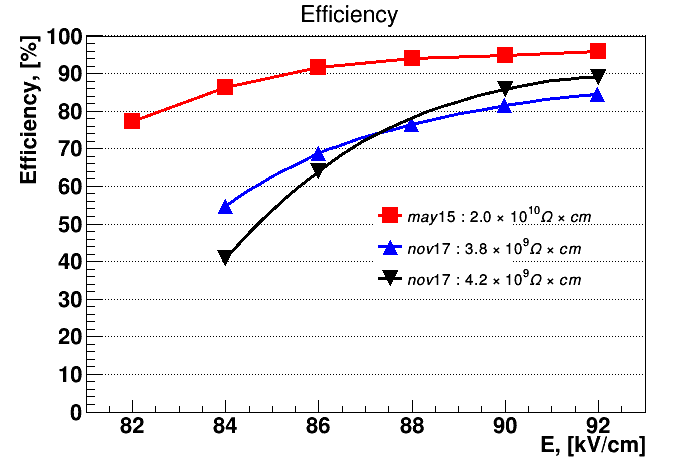} 
\caption{\label{fig:3} RPC efficiency as function of the field strength for three RPC detectors performed 
with two readout systems: squares - MAXIM3760+CAEN TDC, triangles - PADI-10. The corresponding bulk 
resistivities of the ceramic plates on floating potential are given in the legend.}
\label{l2ea4-f3}
\end{figure}

The detector efficiency in dependence on the high voltage studied with the PADI-10 FEE is shown 
in figure~\ref{fig:3} and compared with former measurements. 
One can see that both setups with PADI read-out  do not exhibit a working plateau, reaching 
at maximum 89\,\% registration efficiency. In previous measurements with MAXIM3760 preamplifiers a  wider 
efficiency plateau with a maximum value of 97\,\% has been obtained. 

\begin{figure}[htb]
\centering
\includegraphics*[width=140mm]{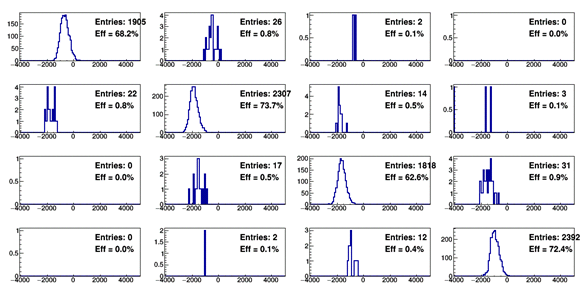} 
\caption{\label{fig:4}Selected time spectra for the central RPC cells \#\,3 - 6 of the BFTC-minimodule. 
In each time spectrum the number of entries and the cell efficiency are shown when the beam hits 
the cell with the highest efficiency in the horizontal line (from cells \#\,3 in the upper line to cells \#\,6 in
the lower line).
}
\label{l2ea4-f4}
\end{figure}

For cross-talk probability determination the BFTC-minimodule cells were placed one-by-one in the beam 
center with a narrow trigger while the signals were read out from all eight RPC cells. A maximum cross-talk 
probability of 1.2\,\%  has been obtained for all other seven channels. This measurement is affected by the 
fact, 
that the beam diameter of 5\,cm (FWHM) is larger than the cell size of 2\,cm. Thus, there is a chance that a 
soft electron, after series of rescatterings, hits both neighboring channels and still passes the trigger 
condition. In a selected number of time spectra of RPC cells located in the center of 
the BFTC-minimodule.  In figure~\ref{fig:4} a selected number of time spectra are shown for  RPC cells 
located in the center of the BFTC-minimodule.  The comparision of the efficiencies in the spectra of horizontal
rows shows, that the suppression of the cross-talk between 
the cell under beam exposure and the neighboring cells is clearly evident.

\begin{figure}[htb]
\centering
\includegraphics*[width=120mm]{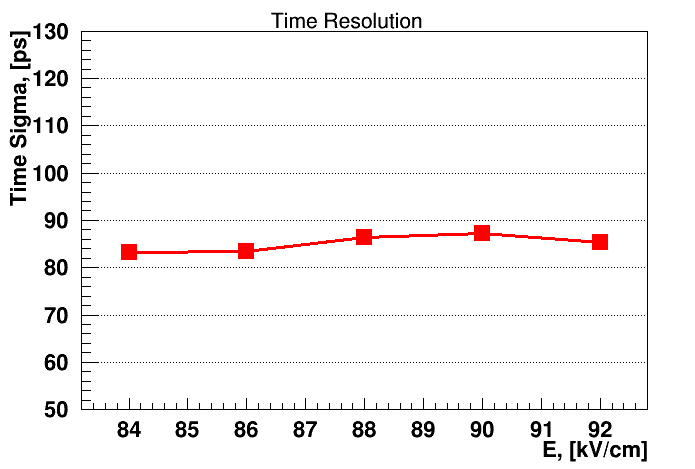} 
\caption{\label{fig:5}Time resolution of cell \#\,3 (bulk resistivity 3.8$\cdot$10$^9 \Omega\cdot$ cm) 
as function of the applied electric field.}
\label{l2ea4-f5}
\end{figure}

A typical time resolution of the RPC as function of the applied electric field is shown in
figure~\ref{fig:5}. 
After calibration of the VFTX TDC the RPC time was corrected for time-walk effect and fitted with 
a Gaussian. The start time resolution of 80\,ps and electronic jitter of 90\,ps the 
RF signal  used as reference  were quadratically subtracted from the width of the Gaussian fit.

\section{Conclusion}
\label{sec:intro}
The BFTC-minimodule of the current design of the inner board demonstrated a stable operation 
together with the PADI-10 preamplifier: no excitation of the readout chain or refracted signals 
in the RPCs time spectra's were observed, the ToT spectra indiscrete with proper shape 
and, the cross-talk probability is very low. This means that the design can be fixed and 
used for future modules, as well as for the (already started) development of the module-prototype 
of size 20$\times$20\,cm$^{2}$ for a test run with heavy ion beam at the upgraded SIS18 at GSI/FAIR. 
The most probable reason for the low registration 
efficiency is related to the short signals which are not correctly processed by the VFTX TDC. 
To check whether this is the case, an external signal stretcher and other preamplifiers will be tested under 
high rates  as well as with cosmic rays. 
\

\end{document}